# SENSOR DATA VALIDATION FOR GARBAGE COLLECTION USING MACHINE LEARNING

●Kabeer Gulati  ●Zuhaib Ahmed  ●Abhishek Raj

(19BEC1023)   (19BEC1399)    ( 19BEC1276)

**Abstract:** Any complex dynamic system's ability to function successfully depends in significant part on the accuracy of the sensor data; hence sensor data validation is crucial. Because sensor data is utilised for monitoring and oversight, erroneous sensor data would result in overall poor process output. In this study, the data-confidence of the sensor data is ascertained using a Mamdani fuzzy inference system. Erroneous data can be corrected with this method. If the sensor outputs faulty value for a prolonged period of time, the system will be reported and a report will be generated. This can be used as a generic module for any system. This fuzzy system is then used upon the readings from an ultrasonic sensor and is used as a part of a bigger and more complex IoT system.

## 1. Introduction:

The performance and safety of a complex system with multiple input sensors are primarily dependent on the quality and consistency of the sensors, as sensor data is responsible for controlling the whole process in the form of control loops. The most common causes of sensor data inaccuracy include age and wear, sample contamination, catalytic degradation of the sample, thermal conductivity, environmental errors such as vibration, heat, turbulence induced by gas flow, and so on. These are only a few of the known causes of sensor data inaccuracy. However, there might be a variety of unknown causes for the incidence of errors. In the majority of mistake occurrences, the operation is momentarily halted and the sensor is replaced. This will result in significant losses for the process industry. As a result, a technique for detecting and correcting sensor errors is required. Sensor data validation refers to the process of determining if the data generated by the sensor is incorrect or correct.

## 2. Literature Review:

In [1] the objective was to have a pragmatic sensor validation system instead of one that focuses on a particular sensor(s) that isn't built into the sensor to reduce power consumption, compute power and cost. The authors used supervised machine learning model to create an error detection layer and used regression based interpolation to replace the erroneous data. They concluded with a process flow for sensor data validation and that the sensor failure would follow a Weibull Distribution curve. In [2] the objective was to have a novel sensor validation architecture to perform SFDIA (Sensor Fault Detection Integration and Accommodation) for air quality sensors and validate the data against publicly available sensor data set. The authors used sensor estimators based on neural networks are constructed for each sensor node in order to accommodate faulty measurements along with a classifier to determine the failure detection and isolation. Results illustrate the prompt detection, isolation and accommodation of sensors' failures with less than

2.6% of faults on average remained undetected on the test set. In [3] the objective was to propose data reliability enhancement method through data schema based data validation and sensing client validation in crowd sensing applications to prevent damage from cyber attacks. The author used data validation based on the data rules desired by the data requester The author concluded that the validation accuracy problem of the existing non data element based validation is solved. The performance evaluation of the proposed method shows that it has the effect of delaying and blocking automated false data attacks using the program.

In[4] the objective was to present a practical approach to automatically validate data from all building's sensors. They designed and implemented four different tests to detect out-of-range values, spikes, latency issues and non-monotonous values for $CO_2$ and temperature sensors. The authors used a combination of Neural Networks and Fuzzy Logic to validate sensor data. They concluded with 2 types of errors; biased values and high latency and emphasised on the importance of data validation and how it affected the building. [5] Aims at detecting anomalies in measurements from sensors, identifying the faulty ones and accommodating them with appropriate estimated data, thus paving the way to reliable digital twins. The authors used general machine learning-based architecture for sensor validation built upon a series of neural-network estimators and a classifier. Estimators correspond to virtual sensors of all unreliable sensors (to reconstruct normal behaviour and replace the isolated faulty sensor within the system), whereas the classifier is used for detection and isolation tasks. They concluded with a three-stage SFDIA architecture with capability to adapt with different applications. In [6] the objective was to assure data quality for perfect decision making in an IoT based Wireless network. The authors use a novel adaptive threshold based approach instead of the machine learning models that we saw more commonly. This approach was more malleable and easy to implement in real time IoT based systems. The prototype/implementation issue was also addressed. From the results, the proposed algorithm appears to be more adaptive, and easy to implement in real-time applications. Moreover, the proposed algorithm has been compared with a recent work DVA and showed a better performance in terms of false detection, delay, and energy consumption.

In [7] the objective was to narrate the handling of a larger number of data generated from multiple sensors and addressing the problems in data fusion for sensitive IoT based applications. In addition to that comparative analysis of various sensor fusion and validation, techniques are surveyed. The authors used a layered architectural framework and guidelines for modeling multi sensor-based applications based on sensor fusion. The advantages of Multisensor data fusion leads to improved Data accuracy, availability and reduces data uncertainty. In [8] the authors aim at the validation of an target acceleration sensor on a truss structure by using Naive Bayesian Classifier and Tree Augmented Naive Bayesian Classifier which are based on machine learning technology whose theory basis is probability statistics. They concluded that that binning number has effect on the classification correct rate. When the binning number increases, the classification accuracy has a downward trend. In [9]the objective is to enhance sensor data security in electricity grid systems. The authors explore the feasibility of using blockchain technology to validate if sensor data is following a known model to increase safety and security in an electrical grid. They concluded that the additional calculations to be negligible in terms of computation time all the while increasing the security of the process. In [10] the objective is to increase reliability, and validation of data collected in distributed edge sensor systems and prevent them from potential cyber attacks. Introducing validation with existing sensors may impose too high a requirement for bandwidth to use cloud-based validation, while edge-based validation may require too much computing power. A fog-based validation layer using sensory substitution is presented. These findings demonstrate that the approach is a viable tool for sensor validation employing accessible sensors, as well as instances in which sensory replacement fixes erroneous positives and erroneous negatives.

In [11] the objective is to make IoT systems more efficient and reliable when used along with sensors that are constantly being operated in real-time by operating each sensor only when we need it to instead of running all the time. The authors propose A framework for sensor correlation and sensor selection to activate and validate a reported anomaly in the monitored environment. They frame and solve sensor selection as a multi objective optimisation (MOO) problem, taking into account sensor power consumption, network quality, remaining battery level, sensor correlation value, and free CPU and RAM of the edge gateway device. The suggested framework utilises minimum edge gateway resources, according to implementation findings.

### 3.  Methodology:

Figure 1 is a generic block diagram for regulating any operation. The process output is sent back into the loop and compared to the set point, and the resulting error is fed into a controller, which changes the final control element to obtain the desired process output. This is only possible if the feedback loop's measuring device (sensor) produces accurate and correct findings. Otherwise, the total process outcome will be flawed.

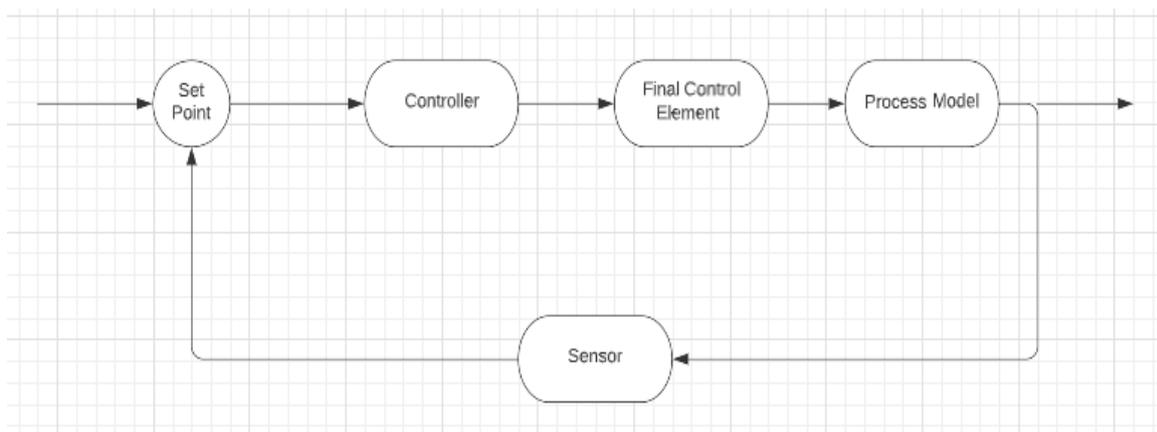

Figure 1.

Figure 2 is the suggested control loop for sensor data validation. Now the process output is validated and reconstructed based on it's accuracy and then is sent to be compared to the set point knowing the data is now accurate and precise because of undergoing Fuzzy based validation.

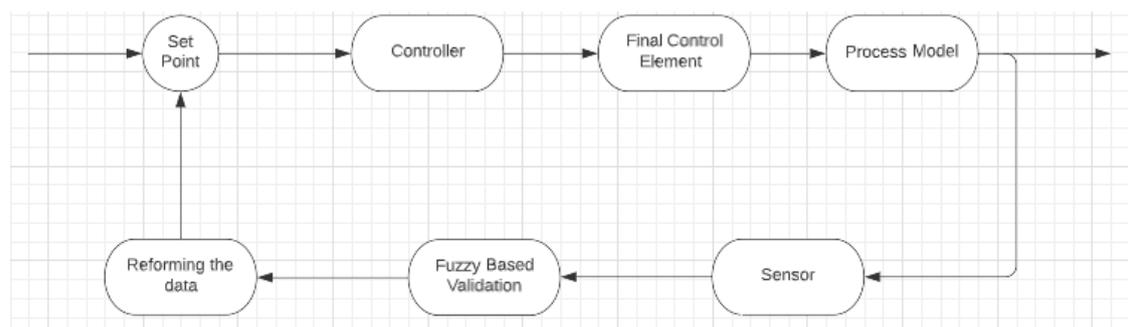

Figure 2.

### 3.1. Error Detection.

For Error Detection we made use of variance of data, uncertainty and principle component analysis (PCA).

### 3.1.1. Variance Based Detection

We can set a threshold variance and if the input signal from the sensors passes that threshold then we can alert the system of a faulty sensor. Here we simulated this by using a random variable and added White Gaussian Noise to it.

### 3.1.2. Uncertainty Based Detection

In this method we calculate the uncertainty according to the Guide to the Expression of Uncertainty(GUM) of the input signal to detect a faulty value. We simulated this by using a random variable with White Gaussian Noise to represent a signal. We then added more noise in the latter half of the input signal to get a distinct output

### 3.1.3. Principle Component Analysis

We can use sensor fusion along with PCA to extract the sensors having higher co-variance and calculate Square Prediction Error (SPE) for only those sensors. Hence reducing computational complexity when working with a sensor fusion (group of sensors that directly or indirectly measure the same stimulus). We simulated this scenario by using two random variables to represent a sensor fusion and later we added noise to one of the sensor for better representation of our use case.

### 3.1.4. Fuzzy Inference System:

**Fuzzy Logic** is a way to model logical reasoning where the truth of a statement is not binary but somewhere between Absolute Truth and Absolute False. **Fuzzy Inference System** is a function that maps input to output using Human Interpretable Rules instead of abstract Mathematics.

### 3.2. Using Fuzzy System for Sensor Data Validation

To implement fuzzy system for sensor data validation we take 3 parameters in consideration. Data from the sensor, Rate of Change (ROC) of data and Standard Deviation

These are the crisp inputs for our fuzzy inference system. Then we plot a membership function for each input; here we used a Gaussian relationship function for illustration purposes. Now we form a relationship matrix that will fuzify our inputs. The output we get from evaluating the fuzzy system will be our data confidence. Now that we have an understanding of what we went ahead to MATLAB's Fuzzy Logic Designer to create a robust fuzzy system that could be implemented cross-platform.

### 3.2.1. Fuzzy Logic Designer

MATLAB offers a Fuzzy Logic Designer that makes it easy and less prone to error to design complex type-1 fuzzy systems with multiple rules and crisp inputs .Here we used the Fuzzy Logic Designer to re-construct our Fuzzy System and make it more robust.

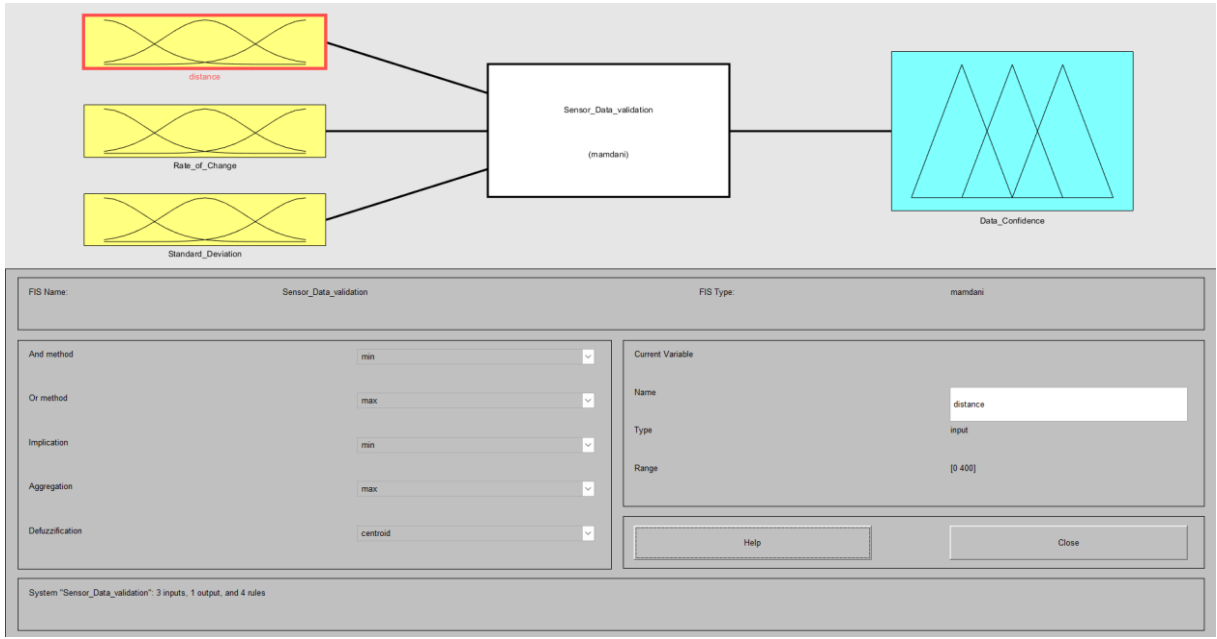

Figure 3.

In Figure 3. we can see the Fuzzy Logic Designer. We have three inputs namely distance, Rate_of_Change and Standard_Deviation. Each of these three crisp inputs have a unique membership. In Figure 6. we can also see the output membership function which decides the data confidence which is what we are aiming to obtain after defuzification.

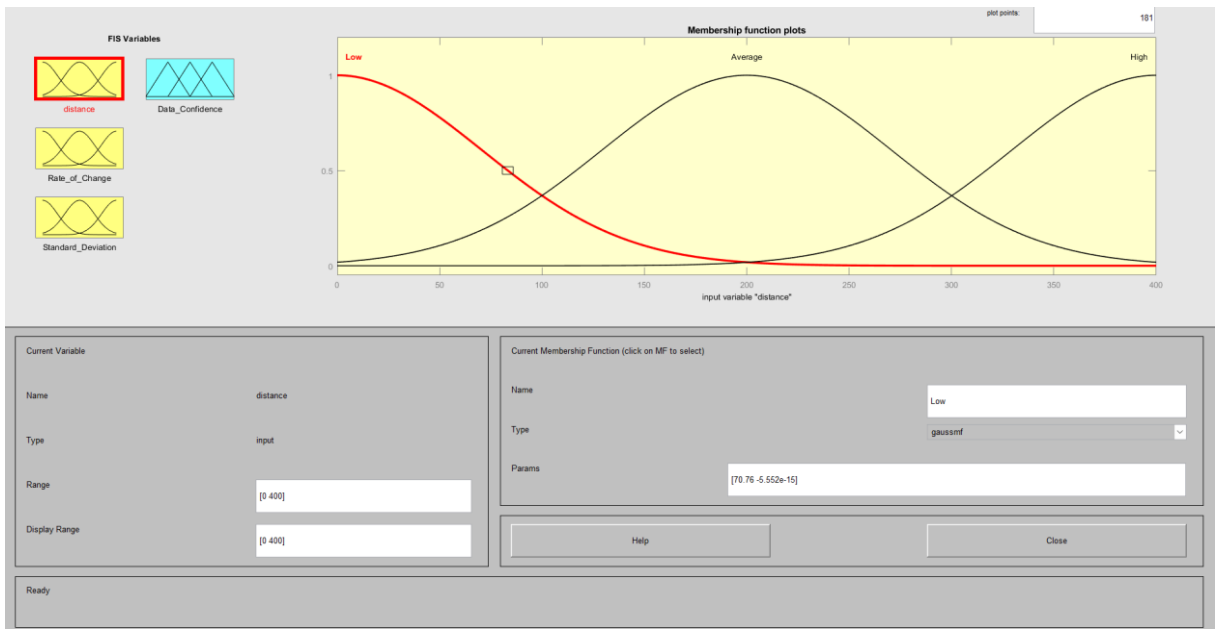

Figure 4.

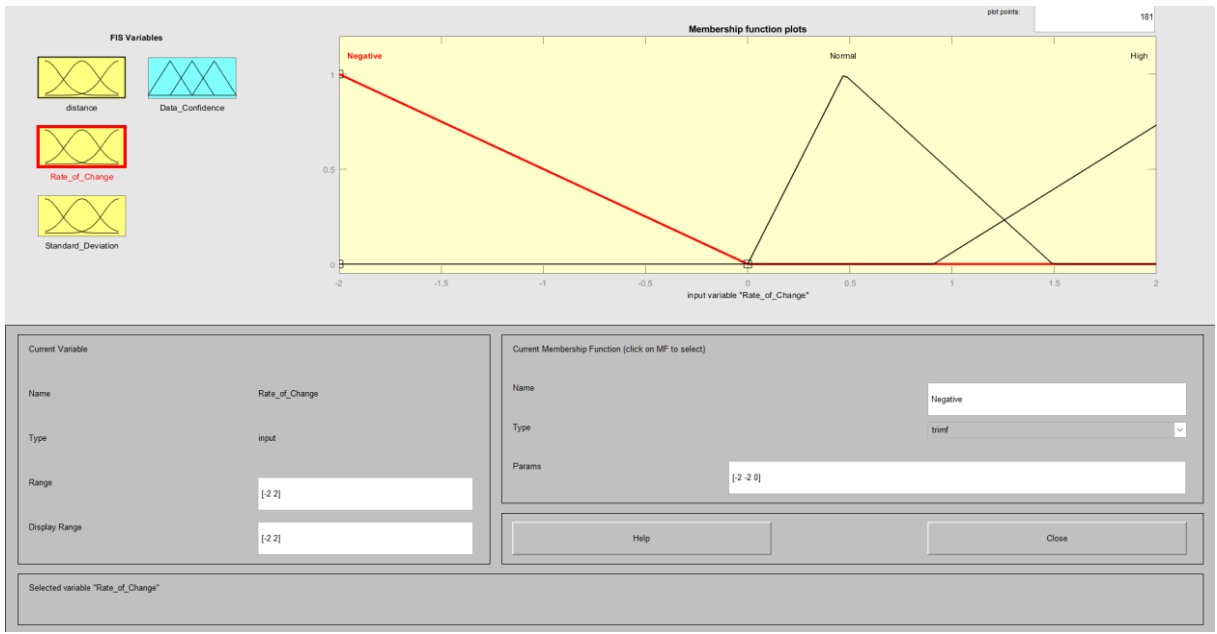

Figure 5.

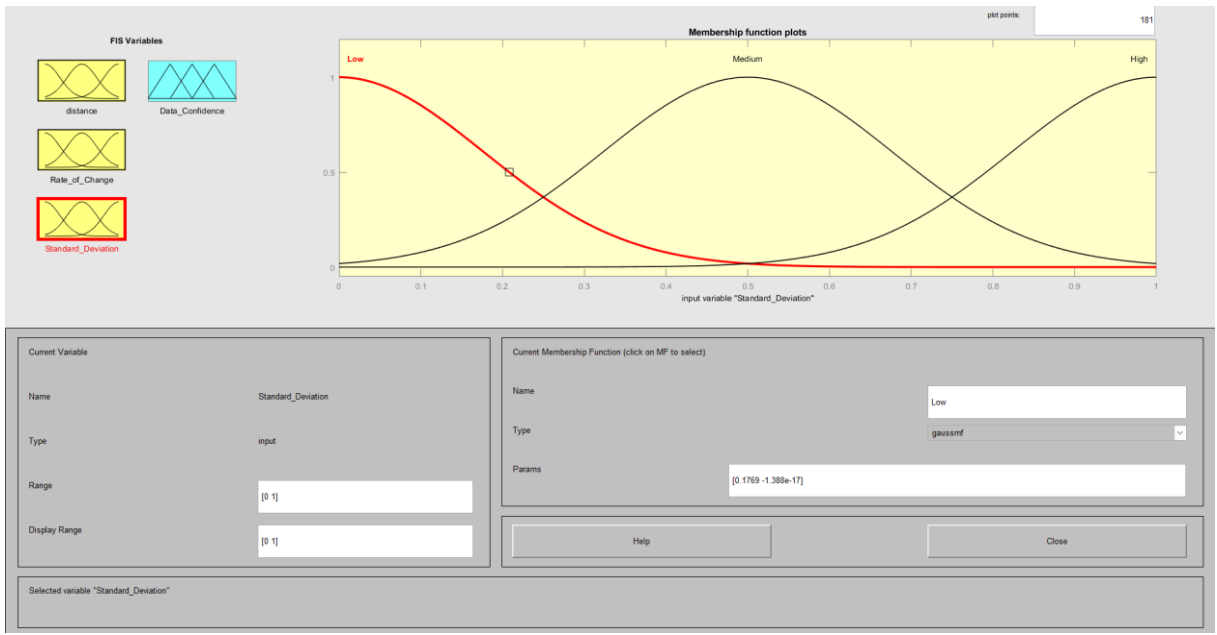

Figure 6.

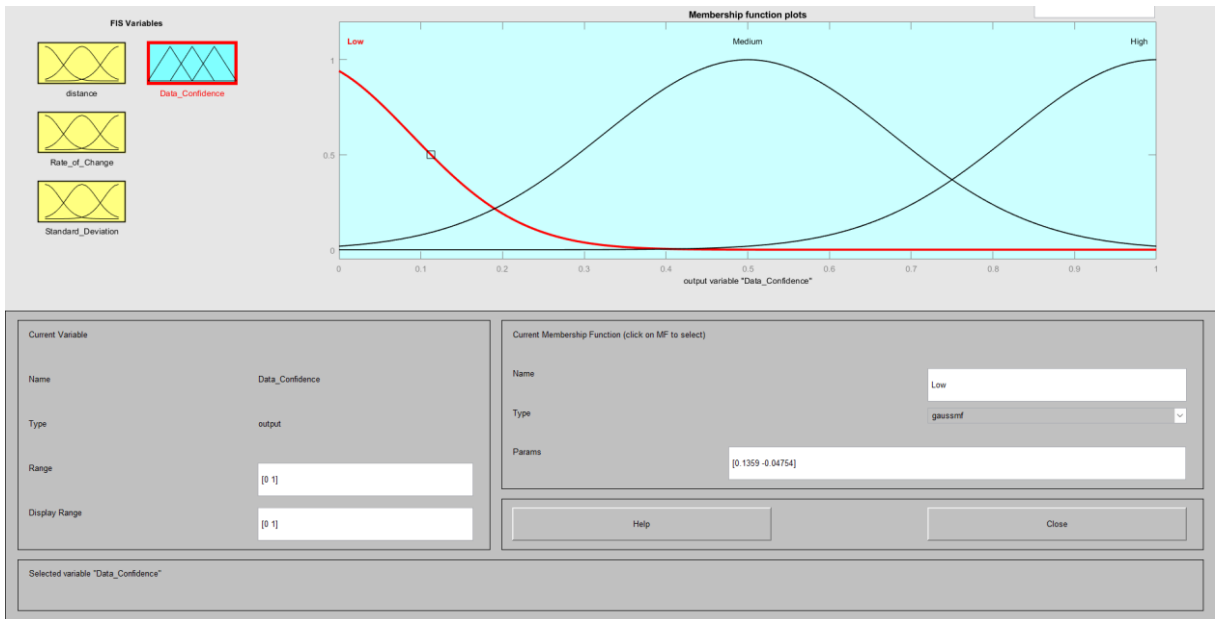

Figure 7.

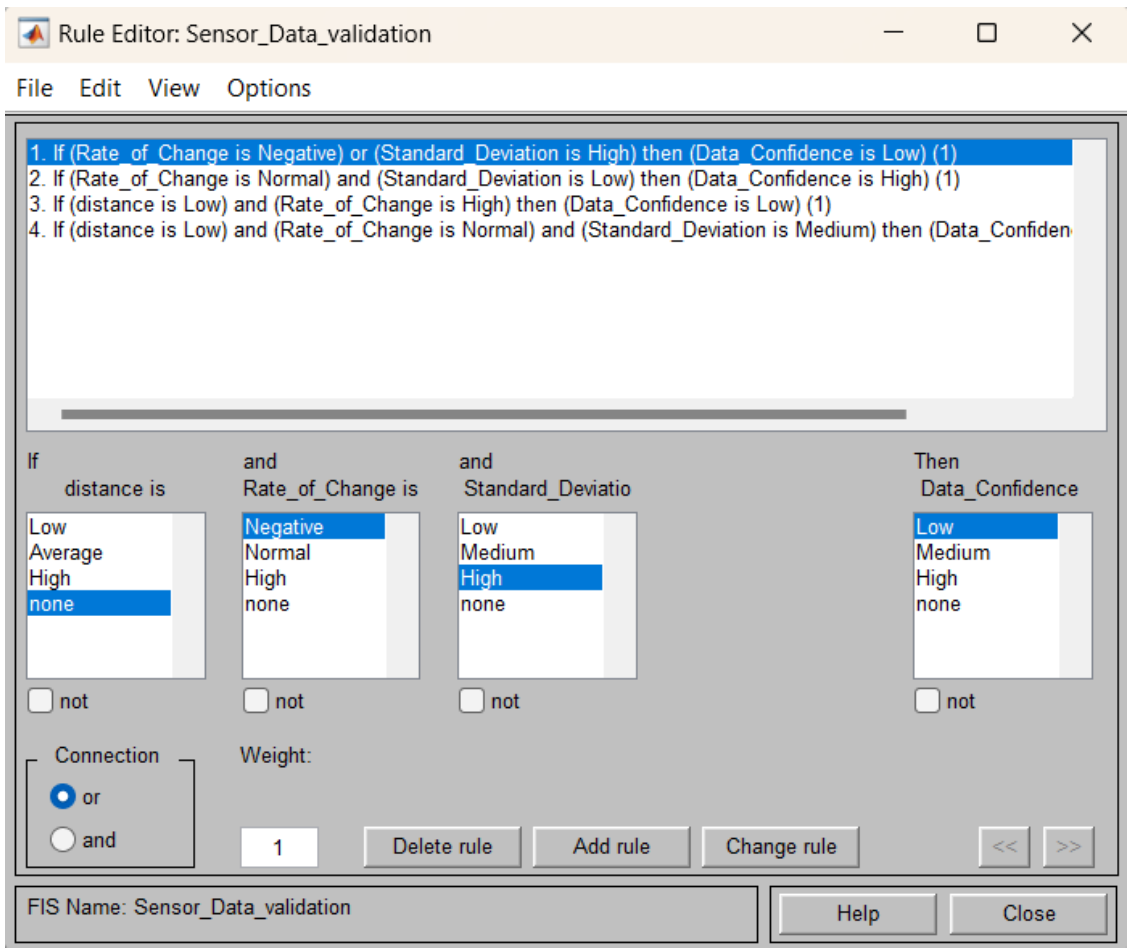

Figure 8.

Figure 8 is the rule editor for our fuzzy system where we can edit and make new rules for our fuzzy system. This interface makes it easier to create new rules as every rule is typed out as a sentence. When we create a fuzzy system in command line we have to define a matrix for rule set which is a complicated task. One more benefit of using the Fuzzy Logic Designer is that it exports the fuzzy system in a .fis file which could be then converted to .c and could be implemented directly on the hardware level.

### 4. Issues:

The Fuzzy Logic Toolbox is not available on Thingspeak hence implementing the fuzzy system online will require converting .fis file to .m . After extensive research we conclude that there is no algorithm out there that could do this reliably.

### 5. Result and Discussion:

Uncertainity Based Detection: In Figure 9 the histogram depicts the dataset that we obtain from our sensor. To display how this method works we added White Gausian Noise after 60 samples. Hence we see an abnormal increase in the Uncertainity Index.

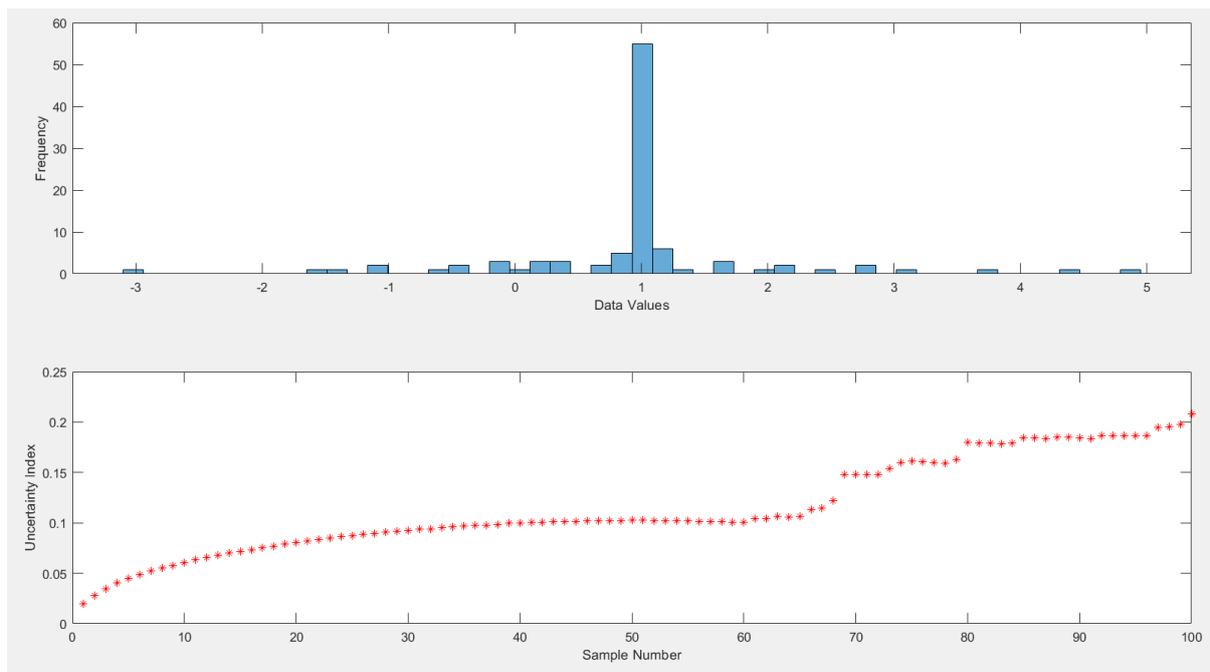

Figure 9.

PCA: Principle Component Analysis was used to study the sensor validation process in a sensor fusion. Figure 10. shows the Data Set and the Normalised Dataset from two sensors a and b when plotted against each other. In Figure 11 we see the Principle Components plotted against each other. The objective of performing PCA is to calculate the Squared Predicted Error. For demonstration purposes we added artificial white gaussain noise to the inputs. In Figure 12 it can clearly be seen how the SPE reacts to that noise.

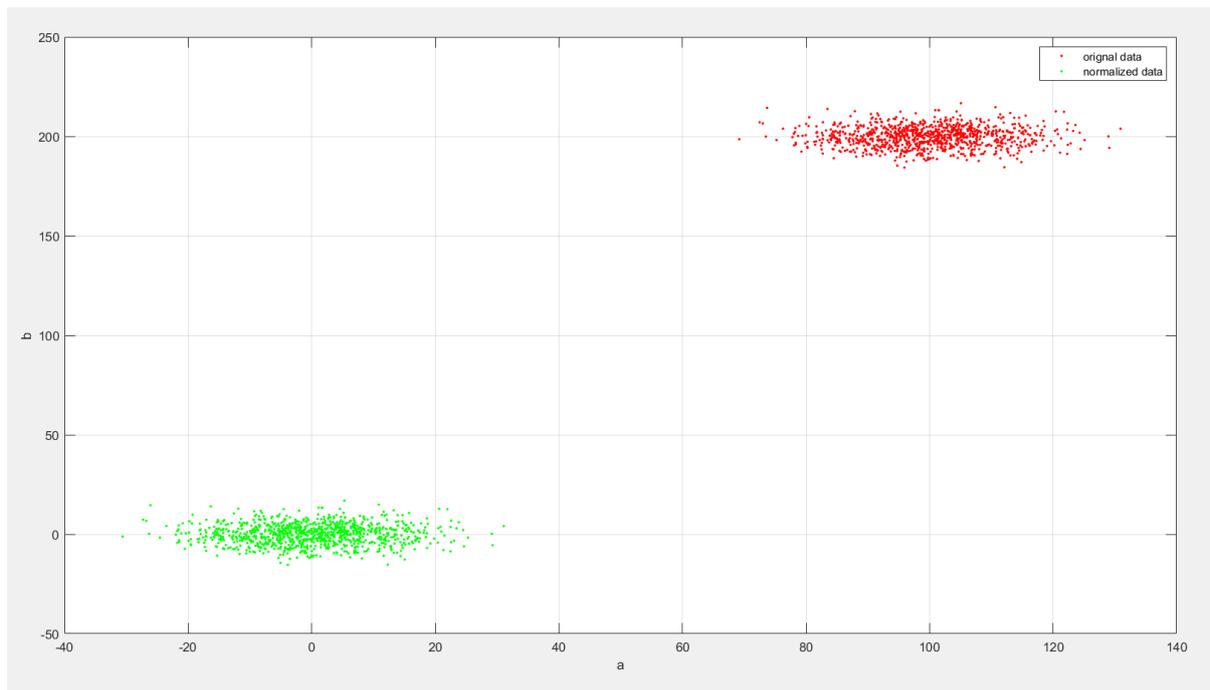

Figure 10.

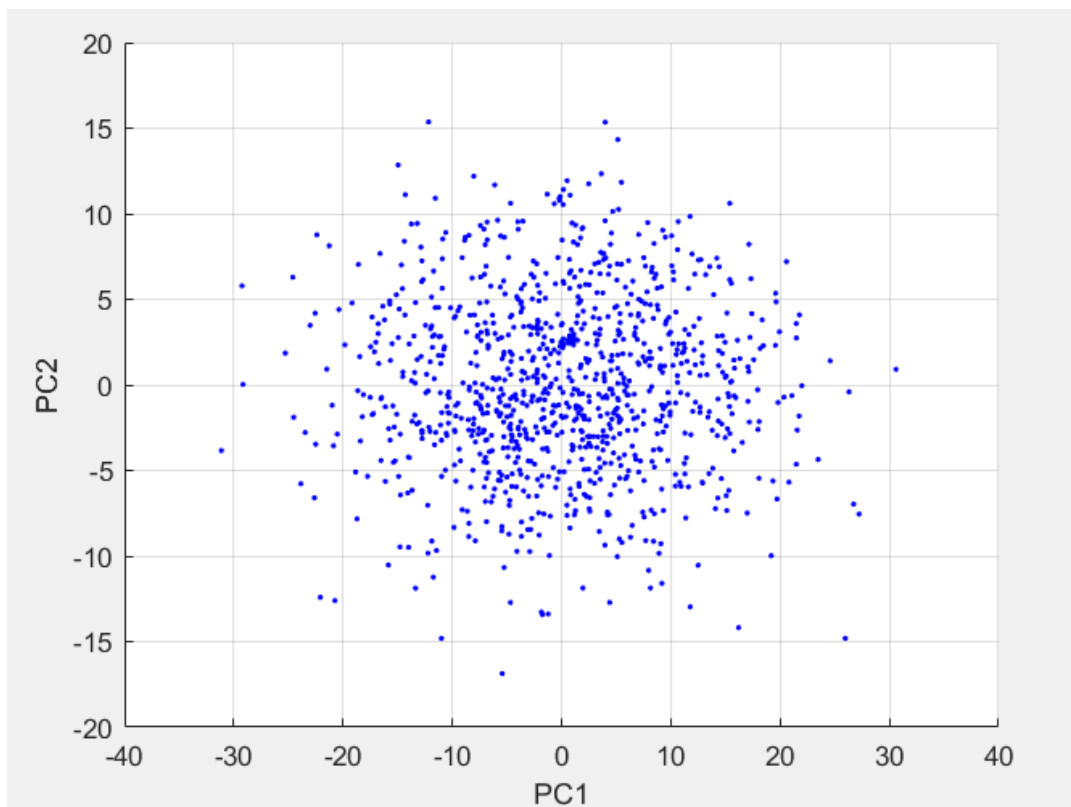

Figure 11.

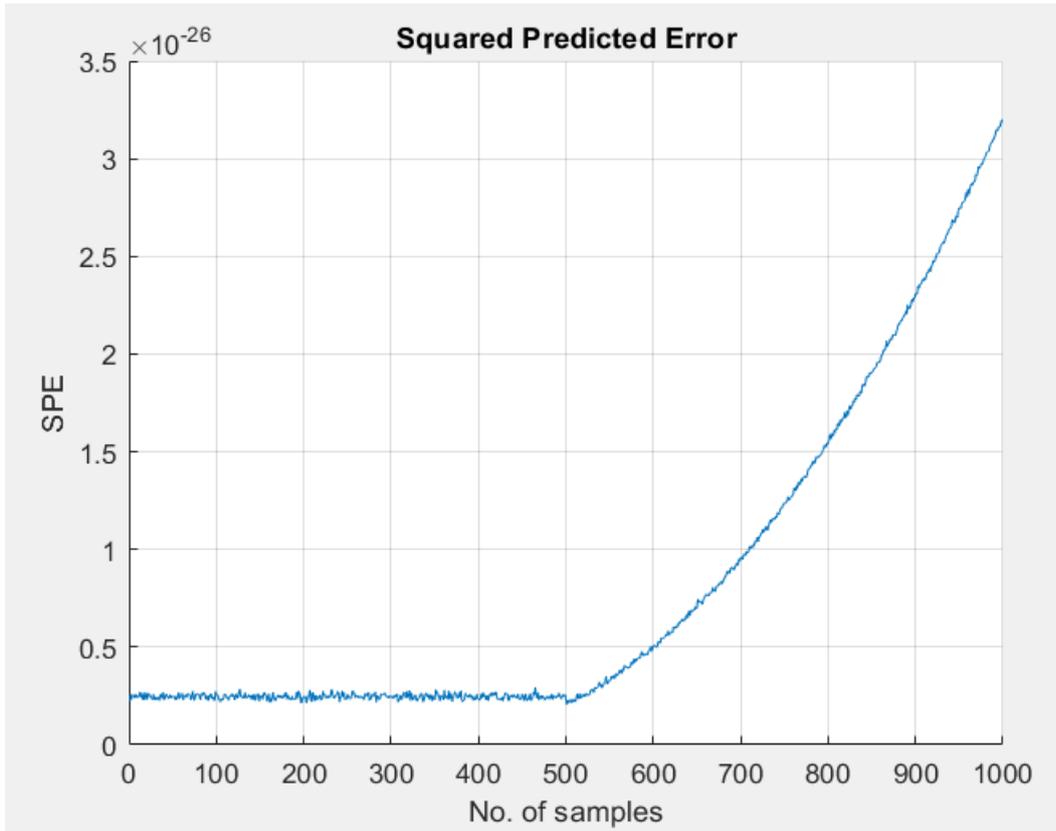

Figure 12.

Fuzzy Logic Designer: From fuzzy logic designer we can obtain the multiple surfaces that represents the relationship between output and various inputs. As we have more than two inputs we have to generate multiple surfaces in order to fully understand the relationship between all of the crisp inputs and output.

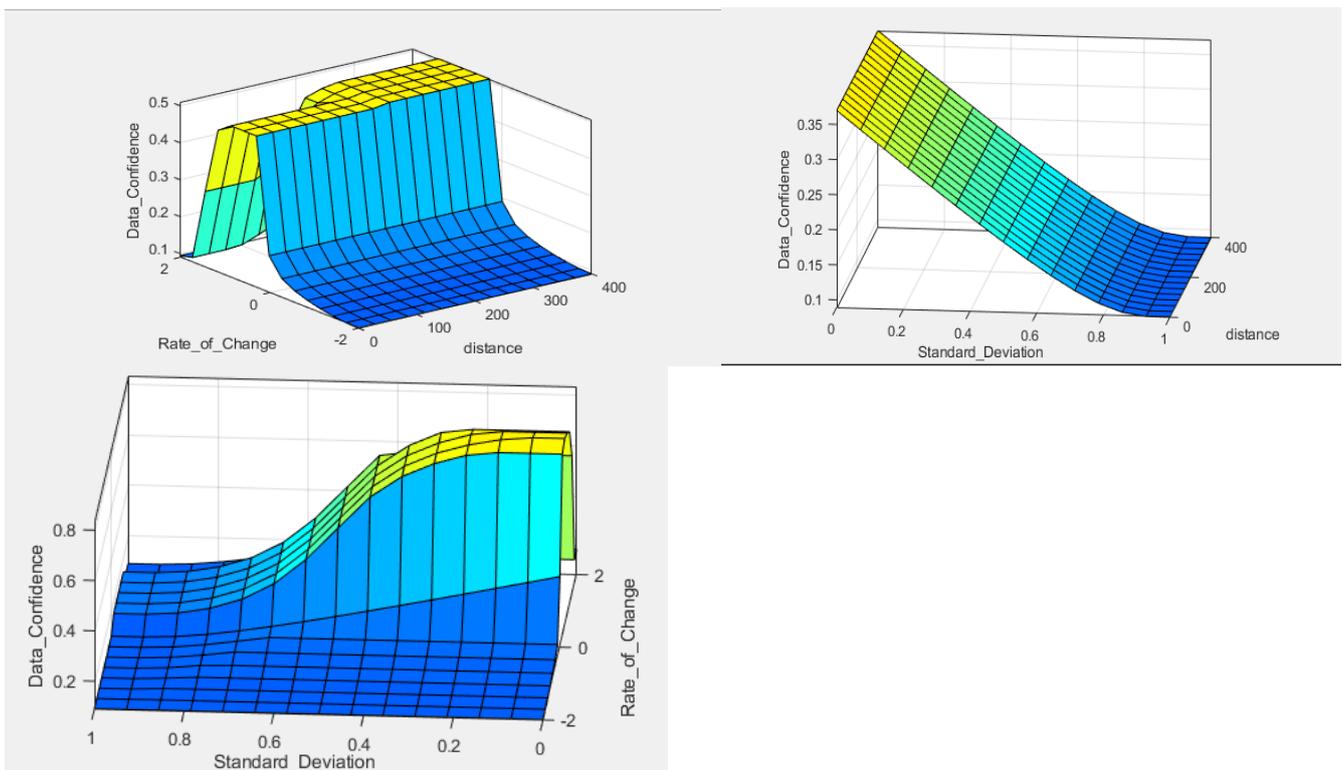